\documentclass[10pt,a4paper,reprint,amsmath,amssymb,aps]{revtex4-1}
\usepackage[latin1]{inputenc}
\usepackage{amsmath}
\usepackage{amsfonts}
\usepackage{amssymb}
\usepackage{graphicx}
\graphicspath{{images/}}

\begin{document}

\title{Entropy of a Zipfian Distributed Lexicon}

\author{Leonardo Carneiro Ara\'{u}jo}
\email{leolca@ufsj.edu.br}
\affiliation{Campus Alto Paraopeba, Universidade Federal de S\~ao Jo\~ao del-Rei,\\ Rod. MG 443, Km 7, Ouro Branco, MG, Brazil}

\author{Tha\"{i}s Crist\'{o}faro-Silva}
\email{thaiscristofarosilva@ufmg.br}
\affiliation{Faculdade de Letras, Universidade Federal de Minas Gerais,\\ Av. Ant\^onio Carlos 6627, 31270-901, Belo Horizonte, MG, Brazil}

\author{Hani Camille Yehia}
\email{hani@ufmg.br}
\affiliation{Departmento de Eletr\^onica, Universidade Federal de Minas Gerais,\\ Av. Ant\^onio Carlos 6627, 31270-901, Belo Horizonte, MG, Brazil}

\date{\today}%

\begin{abstract}
This article presents the calculation of the entropy of a
system with Zipfian distribution and shows that a communication
system tends to present an exponent value close to one, but still
greater than one, so that it might maximize entropy and
hold a feasible lexicon with an increasing size.
This result is in agreement with what is observed in natural
languages and with the balance between the speaker and
listener communication efforts. On the other hand, the
entropy of the communicating source is very sensitive to
the exponent value as well as the length of the observable data,
making it a poor parameter to characterize the communication
process.
\end{abstract}

\maketitle

\section{\label{sec:introduction}Introduction}
Statistical linguistics makes uses of Zipf analysis, which is a statistical
tool used as well in several research fields, such as economics \citep{mandelbrot1963},
gene expression \citep{furusawa}, and chaotic dynamic systems \citep{nicolis1989}.
Zipf found a power-law relation for written texts in
natural languages \citep{zipf1949}. This empirical observation has become the most remarkable
statement in quantitative linguistics. The observation of a Zipf behavior is necessary in
a natural text, as much as it is a necessary behavior of any source
producing information contents, since any randomly generated symbolic sequence
will present a Zipf's law with an exponent between $1$ and $2$ \citep{miller1957,li1992}.
The systematic organization of language reflects the frequencies of usage of
types. Studies have suggested that the frequency of usage is a key factor in
the access of lexical items \citep{balotachumbley1984} and also a driving
force in language change \citep{bybee2002}. Frequency plays an important role
in understanding how human communication works.
A useful example of Zipf analysis is given by \cite{havlin1995}, who suggested
a dissimilarity measure of two Zipf plots, from two different sources,
which will be smaller when the data come from the same source and
larger when they come from different sources. This approach is used
to perform authorship attribution \citep{havlin1995}.

The definitions of entropy and redundancy of a language
were introduced by \cite{shannon1948}. Entropy is
a measure of the average information produced by a source
for each symbol produced in its output. Expressing
the entropy in bits gives us the average number of bits
necessary to express each produced by the source.
Redundancy measures the restrictions imposed on a language
due to its statistical structure, what might be an expression
of physiological and phonological constraints.

The entropy of English printed words was estimated by
\cite{shannon1951} and \cite{grignetti} using a Zipfian
distribution with a characteristic exponent $s=1$.
It is known that natural languages
typically present $s \approx 1$ \citep{piotrovskii}.
Some types of human communications still present
a greater exponent, for example, child speech has been
reported to present $s \approx 1.66$ and military
combat text $s \approx 1.42$.
Studies on animal communication also present a Zipf's
behaviour, for example, \cite{mccowan1999} present an exponent
value of $s \approx 1.1$ and $s \approx 0.87$ for adult
and infant dolphins, respectively.
The value of the exponent $s$ seems to describe the plasticity
of the communication system, what leads to a potentially
growing lexicon. Larger values of $s$ characterize systems still
in formation and small values systems well-grounded.
In this paper, we are going to present
the calculation of the entropy of a system using an arbitrary Zipfian
distribution and verify the effect of the
characteristic exponent $s$ on the entropy of the system.

\section{\label{sec:entropy}Entropy of the System}
The entropy of a system using $N$ symbols of probabilities $p_k$,
where $k=1$ to $N$, is given by
\begin{equation}
\label{eq:entropy}
\overline{H} = - \sum_{k=1}^N p_{k} \log_2 p_{k} = -\frac{1}{\ln 2} \sum_{k=1}^N p_{k} \ln p_{k} \textmd{ .}
\end{equation}
If we consider words as the symbols used by our system,
the probabilities $p_k$ might be estimated by counting the frequency
of occurrence of types and dividing it by our sample size.
No matter how large our corpus is, there might always be words
in the underlying lexicon that have no representation on that
corpus. In order to acquire a better estimate of the probabilities,
we should perform a Turing Smoothing \citep{Good1953,gale1994}.

George Kingsley Zipf made important contributions on
language statistics, performing word count experiments, from which
he determined that there is a relationship between word's frequency of
appearance in texts and its rank, the product of them being roughly a constant
\citep{zipf1949}.
The distribution of words in a language follows a power law:
$p_k(s,N) = C k^{-s}$, where
$p_k$ stands for probability of occurrence of the $k$-th most frequent
word in the corpus; $C$ is a normalizing constant,
$C^{-1}=\sum_{n=1}^N n^{-s}$, which is the generalized harmonic number;
$k$ is the word rank; $s$ the slope, which characterizes the
distribution; and $N$ is the number of elements in the set.

Zipf's law seems to hold regardless the language observed \citep{zipf1949}.
``Investigations with English, Latin, Greek, Dakota, Plains Cree,
Nootka (an Eskimo language), speech of children at various ages,
and some schizophrenic speech have all been seen to follow
this law''\citep{weiss1998}.
Since a smoothing is necessary, in order to achieve a better approximation
of the underlying probabilities, it is important to notice
that there is a relation between
the Turing's smoothing formula and Zipf's law: both are shown to be
instances of a common class of re-estimation formula
and Turing's formula ``smooths the frequency estimates
towards a geometric distribution. (...) Although the two equations
are similar, Turing's formula shifts the frequency mass towards more
frequent species''\citep{samuelsson1996}.

\begin{figure}[htbp]
\centering
\includegraphics[width=0.5\textwidth]{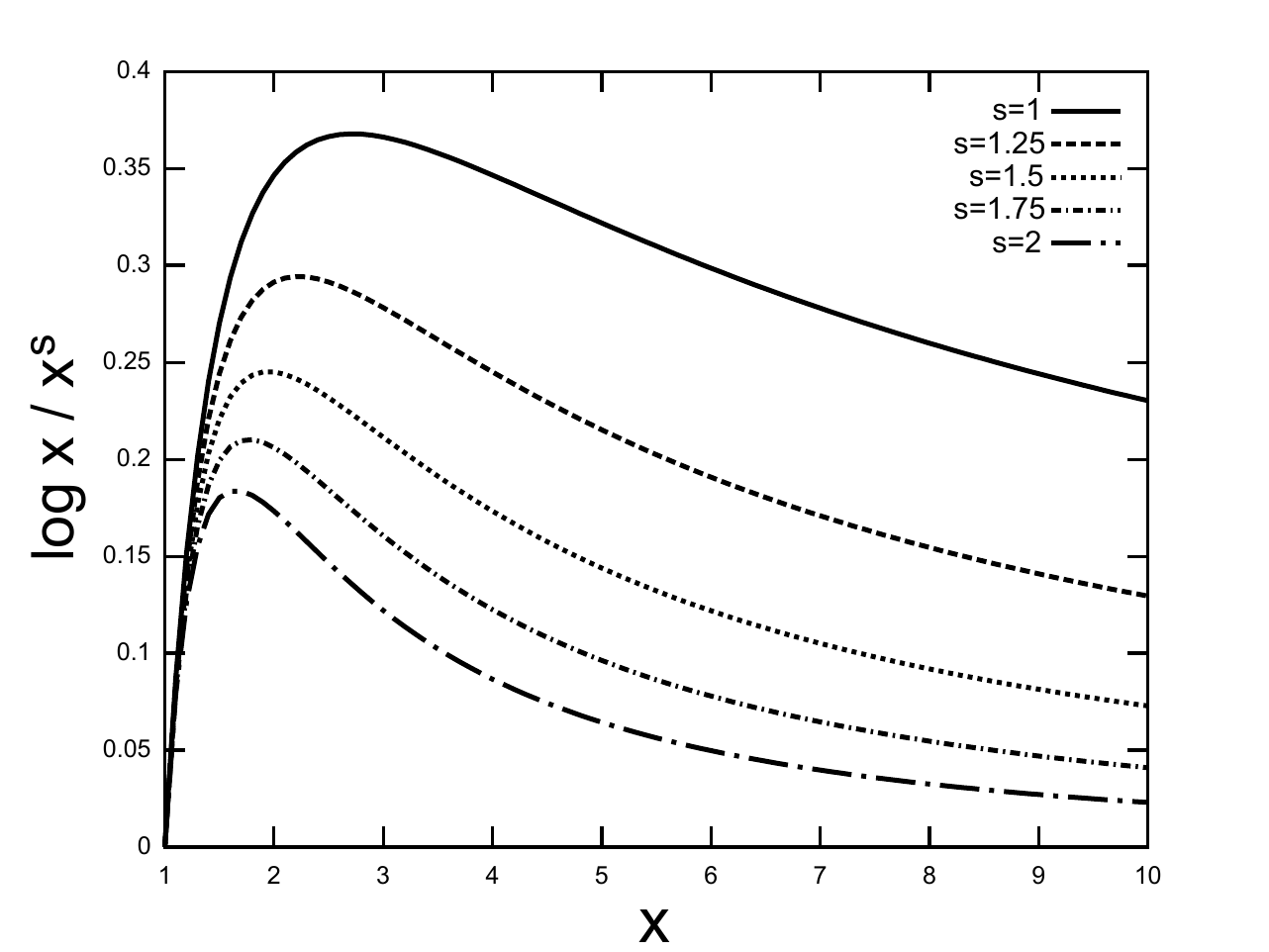}
\caption{Function $f(x)=\ln x / x^s$ for different values of $s$.}
\label{fig:logk_ks}
\end{figure}

Using the Zipfian value for the probabilities in Equation \ref{eq:entropy},
we get
\begin{eqnarray}
\label{eq:ent_zipfian_source}
\overline{H} &=& -\frac{1}{\ln 2} \sum_{k=1}^N Ck^{-s} \ln (Ck^{-s}) \nonumber \\
             &=& \frac{sC}{\ln 2} \sum_{k=1}^N \frac{\ln k}{k^s} - \frac{\ln C}{\ln 2} \textmd{ ,}
\end{eqnarray}
where the summation might be calculated following the steps proposed by
\cite{grignetti}. Figure \ref{fig:logk_ks} presents the function
$f(x)= x^{-s} \ln x $ for different values of $s$ greater than one,
which is usually found in human languages. From the first derivative of $f$,
\begin{equation}
f'(x) = x^{-s-1} (1 - s \ln x) \textmd{ ,}
\end{equation}
we might conclude that $f$ is a decreasing function for $x > e^{1/s}$,
what might be verified in the Figure \ref{fig:logk_ks}.

Particularly for $s \geq 1$, the function $f$
will be a decreasing function for $x>3$.
We might then approximate the summation
using the Riemann sum approximation of an integral. The left Riemann sum
$S_l$ is an overestimate and the right Riemann sum $S_r$ is an underestimate,
\begin{equation}
\label{eq:riemmansumestimate}
S_r \leq \int_{a}^{b} f(x) dx \leq S_l \textmd{ .}
\end{equation}

Using Equation \ref{eq:riemmansumestimate} we might write
\begin{equation}
\label{eq:riemmansume1}
\sum_{n=4}^{N-1} \frac{\ln n}{n^s} \leq \int_{3}^{N-1} \frac{\ln x}{x^s} dx \leq \sum_{n=3}^{N-2} \frac{\ln n}{n^s}
\end{equation}
and
\begin{equation}
\label{eq:riemmansume2}
\sum_{n=4}^{N} \frac{\ln n}{n^s} \leq \int_{3}^{N} \frac{\ln x}{x^s} dx \leq \sum_{n=3}^{N-1} \frac{\ln n}{n^s} \textmd{ .}
\end{equation}
From Equations \ref{eq:riemmansume1} and \ref{eq:riemmansume2} we conclude that
\begin{equation}
\label{eq:riemmansume12}
\int_{3}^{N} \frac{\ln x}{x^s} dx \leq \sum_{n=3}^{N-1} \frac{\ln n}{n^s} \leq \int_{3}^{N-1} \frac{\ln x}{x^s} dx + \frac{\ln 3}{3^s}
\end{equation}
and, by adding the remaining terms (i.e. $n=1$ and $2$) to the summation, we get
\begin{equation}
\label{eq:riemmansume12f}
\int_{3}^{N} \frac{\ln x}{x^s} dx + \frac{\ln 2}{2^s} \leq \sum_{n=1}^{N-1} \frac{\ln n}{n^s} \leq \int_{3}^{N-1} \frac{\ln x}{x^s} dx + \frac{\ln 3}{3^s} + \frac{\ln 2}{2^s} \textmd{ .}
\end{equation}
The wanted bounds are given by adding the $N$-th term to the above equation,
\begin{multline}
\label{eq:riemmansumeboundsfinal}
\int_{3}^{N} \frac{\ln x}{x^s} dx + \frac{\ln 2}{2^s} + \frac{\ln N}{N^s} \leq \sum_{n=1}^{N} \frac{\ln n}{n^s} \\
\leq \int_{3}^{N-1} \frac{\ln x}{x^s} dx + \frac{\ln 3}{3^s} + \frac{\ln 2}{2^s} + \frac{\ln N}{N^s} \textmd{ .}
\end{multline}

\begin{figure}[htbp]
\centering
\includegraphics[width=0.5\textwidth]{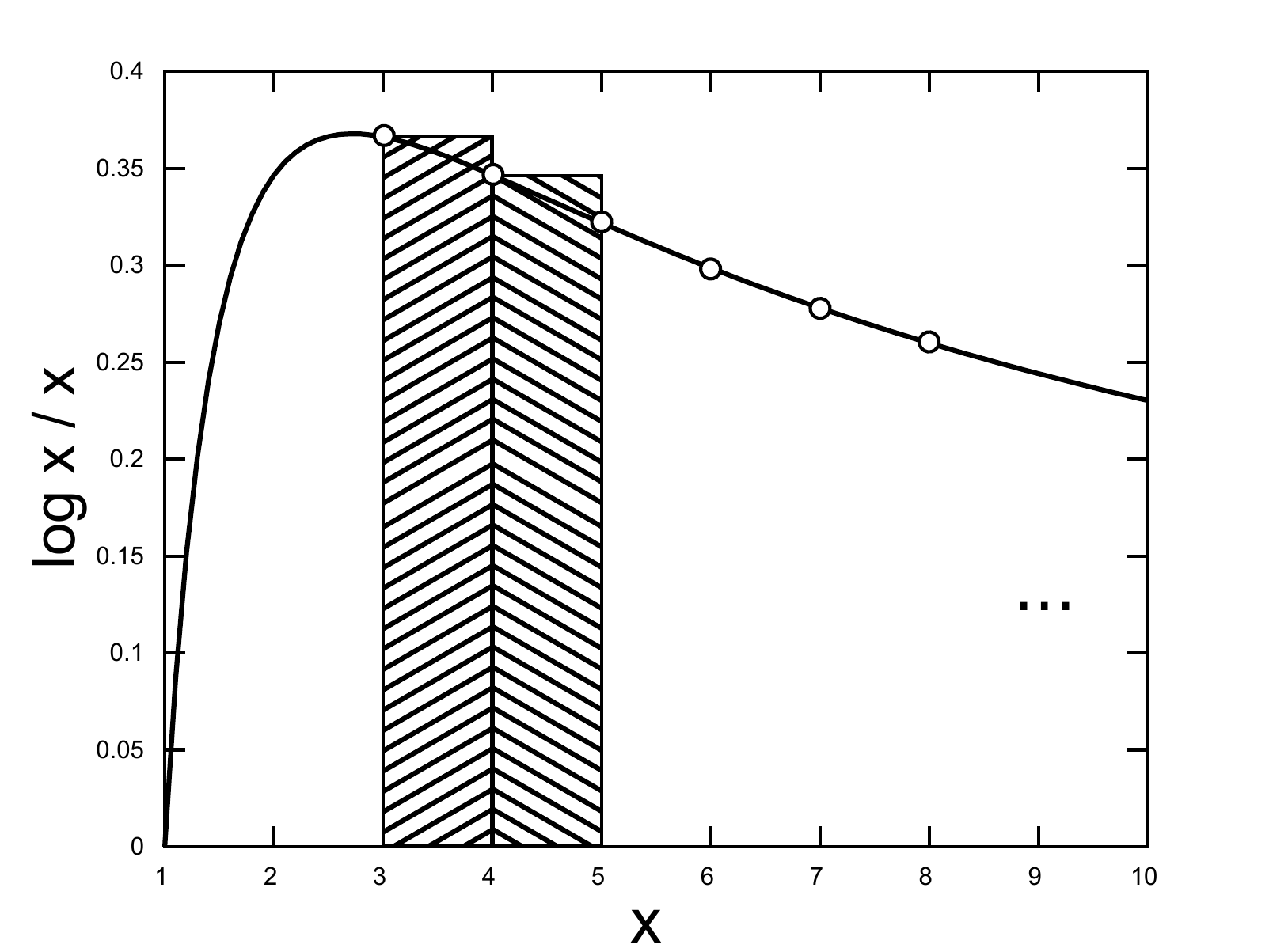}
\caption{Left Riemann sum approximation of the integral.}
\label{fig:logk_k_integral}
\end{figure}

The proposed approximation procedure is illustrated in
Figure \ref{fig:logk_k_integral}.
The integral in Equation \ref{eq:riemmansumeboundsfinal} is solved by parts,
giving the following result, when $s \neq 1$:
\begin{equation}
\label{eq:intparts}
\int \frac{\ln x}{x^s} dx =  \frac{x^{1-s}}{1-s} \left( \ln x - \frac{1}{1-s} \right) \textmd{ ,}
\end{equation}
where the integration constant is omitted, since it is irrelevant when
evaluating the integral in an interval. When $s=1$ the integral will
result in
\begin{equation}
\label{eq:intpartss1}
\int \frac{\ln x}{x} dx = \frac{(\ln x)^2}{2} \textmd{ .}
\end{equation}

\begin{figure}[htbp]
\centering
\includegraphics[width=0.5\textwidth]{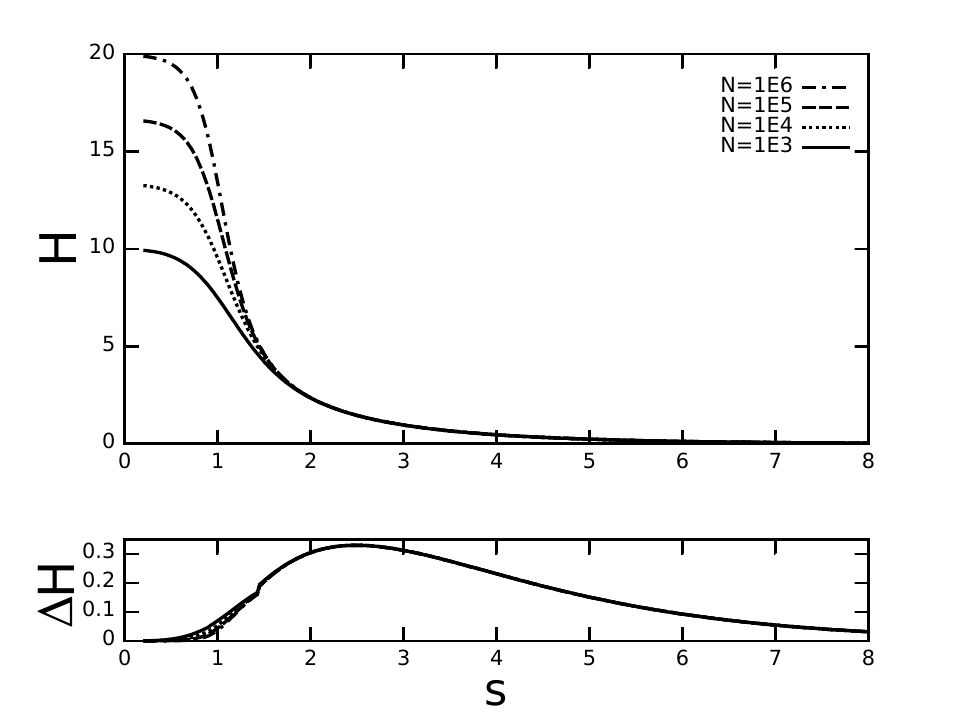}
\caption{Entropy $H$ (in bits) as a function of the Zipf exponent $s$ and sample length $N$. The upper plot presents the average Entropy estimated and the lower plot presents the difference between the upper and lower bounds of the entropy estimated.}
\label{fig:entropy_N_s}
\end{figure}

Using Equations \ref{eq:ent_zipfian_source}, \ref{eq:riemmansumeboundsfinal}
and \ref{eq:intparts} (or \ref{eq:intpartss1}) we are able to calculate the bounds of the
entropy of a Zipfian distributed source for a given $s$ and $N$.
Figure \ref{fig:entropy_N_s} presents some results for different
corpora lengths. We might observe
that the entropy decreases with $s$ and increases with $N$.

\section{\label{sec:conclusion}Conclusion}
The entropy of a system with Zipfian distributed symbols
is decreasing with the characteristic
exponent $s$. A value of $s$ greater than one is a necessary
condition for the convergence of the generalized harmonic number.
In the limit, it is regarded as the Riemann zeta function,
which converges for real $s>1$. An exponent $s$ which satisfies this
condition leads to a Zipfian distribution of the lexicon which
will hold regardless how big the lexicon is.

This limiting value
of $s$ close to one is also found by \cite{ramon2003}
when they proposed ``an energy function combining the effort for the
hearer and the effort for the speaker''.
The minimization of this function leads to a Zipfian distribution
with $s=1$, which is consistent with what is found in human languages.
An exponent $s$ greater than $1$ is necessary to guarantee an hypothetical
growing lexicon without bounds. We might then expect a greater exponent
when language acquisition is still in process and a smaller exponent, closer
to one, when this learning period is consolidating. Rudimentary and
severely restricted communication processes might experience an exponent
smaller than one, since they are not expected to evolve and widen through time,
and that choice maximizes the entropy of the source.

\begin{figure}[htbp]
\centering
\includegraphics[width=0.5\textwidth]{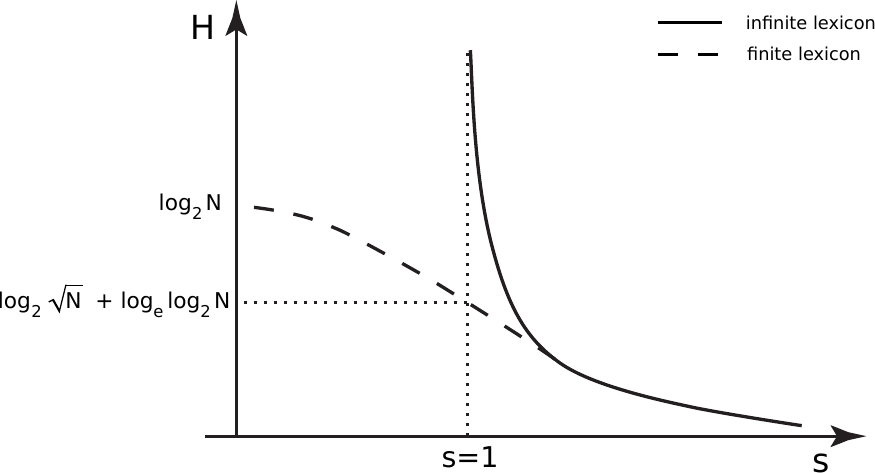}
\caption{The entropy of a source with Zipfian distribution as a function of the characteristic exponent. A finite lexicon and an infinite lexicon behaviours are compared (adapted from \cite{mandelbrot1953}).}
\label{fig:entropy_zipf}
\end{figure}

The maximum rank and the repertoire size are influenced by the length
of our observation but, in practical aspects, it will always be limited
due to our finite observation interval. It will always lead to a finite
lexicon, the set of words observed in the sample.
An infinite lexicon is only a hypothetical approximation, which
is important to analyze under the assumption of the constantly growing
underlying lexicon used in human communications.
Figure \ref{fig:entropy_zipf} presents an adaptation from
\cite{mandelbrot1953}, where the entropy of an finite and infinite
lexicon are compared as a function of the Zipf exponent.
From both figures \ref{fig:entropy_N_s} and \ref{fig:entropy_zipf},
we might observe that the length of the observation sample is
crucial in determining the entropy of the source. A simple truncation
on the sample may lead to a severe distortion on the entropy estimate.
It is also important to note that the entropy estimate is much more
sensitive for $s$ in the vicinity of $1$, meaning that two
sources with different characteristics might have similar
values of their Zipf exponent, but present quite different
entropy estimates.
The exponent $s$ might then be a poor parameter to characterize
the information in a communication process whose symbols
are Zipfian distributed.
Therefore, if the Zipf distribution is used straightforwardly in the
estimation of the entropy of the source, attention must be paid to its
actual meaning

\begin{acknowledgments}
This work has been supported by the Brazilian agencies CNPq and FAPEMIG.
\end{acknowledgments}

\nocite{*}

\bibliography{zipfentropy}

\begin{thebibliography}{24}%
\makeatletter
\providecommand \@ifxundefined [1]{%
 \@ifx{#1\undefined}
}%
\providecommand \@ifnum [1]{%
 \ifnum #1\expandafter \@firstoftwo
 \else \expandafter \@secondoftwo
 \fi
}%
\providecommand \@ifx [1]{%
 \ifx #1\expandafter \@firstoftwo
 \else \expandafter \@secondoftwo
 \fi
}%
\providecommand \natexlab [1]{#1}%
\providecommand \enquote  [1]{``#1''}%
\providecommand \bibnamefont  [1]{#1}%
\providecommand \bibfnamefont [1]{#1}%
\providecommand \citenamefont [1]{#1}%
\providecommand \href@noop [0]{\@secondoftwo}%
\providecommand \href [0]{\begingroup \@sanitize@url \@href}%
\providecommand \@href[1]{\@@startlink{#1}\@@href}%
\providecommand \@@href[1]{\endgroup#1\@@endlink}%
\providecommand \@sanitize@url [0]{\catcode `\\12\catcode `\$12\catcode
  `\&12\catcode `\#12\catcode `\^12\catcode `\_12\catcode `\%12\relax}%
\providecommand \@@startlink[1]{}%
\providecommand \@@endlink[0]{}%
\providecommand \url  [0]{\begingroup\@sanitize@url \@url }%
\providecommand \@url [1]{\endgroup\@href {#1}{\urlprefix }}%
\providecommand \urlprefix  [0]{URL }%
\providecommand \Eprint [0]{\href }%
\providecommand \doibase [0]{http://dx.doi.org/}%
\providecommand \selectlanguage [0]{\@gobble}%
\providecommand \bibinfo  [0]{\@secondoftwo}%
\providecommand \bibfield  [0]{\@secondoftwo}%
\providecommand \translation [1]{[#1]}%
\providecommand \BibitemOpen [0]{}%
\providecommand \bibitemStop [0]{}%
\providecommand \bibitemNoStop [0]{.\EOS\space}%
\providecommand \EOS [0]{\spacefactor3000\relax}%
\providecommand \BibitemShut  [1]{\csname bibitem#1\endcsname}%
\let\auto@bib@innerbib\@empty
\bibitem [{\citenamefont {Mandelbrot}(1963)}]{mandelbrot1963}%
  \BibitemOpen
  \bibfield  {author} {\bibinfo {author} {\bibfnamefont {B.}~\bibnamefont
  {Mandelbrot}},\ }\href@noop {} {\emph {\bibinfo {title} {Oligopoly, Mergers,
  and the Paretian Size Distribution of Firms}}},\ \bibinfo {type} {External
  Research Note: NC-246}\ (\bibinfo  {institution} {IBM},\ \bibinfo {year}
  {1963})\BibitemShut {NoStop}%
\bibitem [{\citenamefont {Furusawa}\ and\ \citenamefont
  {Kaneko}(2003)}]{furusawa}%
  \BibitemOpen
  \bibfield  {author} {\bibinfo {author} {\bibfnamefont {C.}~\bibnamefont
  {Furusawa}}\ and\ \bibinfo {author} {\bibfnamefont {K.}~\bibnamefont
  {Kaneko}},\ }\href@noop {} {\bibfield  {journal} {\bibinfo  {journal}
  {Physical review letters}\ }\textbf {\bibinfo {volume} {90}} (\bibinfo {year}
  {2003})}\BibitemShut {NoStop}%
\bibitem [{\citenamefont {Nicolis}\ \emph {et~al.}(1989)\citenamefont
  {Nicolis}, \citenamefont {Nicolis},\ and\ \citenamefont
  {Nicolis}}]{nicolis1989}%
  \BibitemOpen
  \bibfield  {author} {\bibinfo {author} {\bibfnamefont {G.}~\bibnamefont
  {Nicolis}}, \bibinfo {author} {\bibfnamefont {C.}~\bibnamefont {Nicolis}}, \
  and\ \bibinfo {author} {\bibfnamefont {J.~S.}\ \bibnamefont {Nicolis}},\
  }\href@noop {} {\bibfield  {journal} {\bibinfo  {journal} {Journal of
  Statistical Physics}\ }\textbf {\bibinfo {volume} {54}},\ \bibinfo {pages}
  {915} (\bibinfo {year} {1989})}\BibitemShut {NoStop}%
\bibitem [{\citenamefont {Zipf}(1949)}]{zipf1949}%
  \BibitemOpen
  \bibfield  {author} {\bibinfo {author} {\bibfnamefont {G.~K.}\ \bibnamefont
  {Zipf}},\ }\href@noop {} {\emph {\bibinfo {title} {Human Behaviour and the
  Principle of Least Effort: An Introduction to Human Ecology}}}\ (\bibinfo
  {publisher} {Hafner Pub. Co},\ \bibinfo {year} {1949})\BibitemShut {NoStop}%
\bibitem [{\citenamefont {Miller}(1957)}]{miller1957}%
  \BibitemOpen
  \bibfield  {author} {\bibinfo {author} {\bibfnamefont {G.~A.}\ \bibnamefont
  {Miller}},\ }\href@noop {} {\bibfield  {journal} {\bibinfo  {journal} {The
  American Journal of Psychology}\ }\textbf {\bibinfo {volume} {70}},\ \bibinfo
  {pages} {311} (\bibinfo {year} {1957})}\BibitemShut {NoStop}%
\bibitem [{\citenamefont {Li}(1992)}]{li1992}%
  \BibitemOpen
  \bibfield  {author} {\bibinfo {author} {\bibfnamefont {W.}~\bibnamefont
  {Li}},\ }\href@noop {} {\bibfield  {journal} {\bibinfo  {journal} {IEEE
  Transactions on Information Theory}\ }\textbf {\bibinfo {volume} {38}},\
  \bibinfo {pages} {1842} (\bibinfo {year} {1992})}\BibitemShut {NoStop}%
\bibitem [{\citenamefont {Balota}\ and\ \citenamefont
  {Chumbley}(1984)}]{balotachumbley1984}%
  \BibitemOpen
  \bibfield  {author} {\bibinfo {author} {\bibfnamefont {D.~A.}\ \bibnamefont
  {Balota}}\ and\ \bibinfo {author} {\bibfnamefont {J.~I.}\ \bibnamefont
  {Chumbley}},\ }\href@noop {} {\bibfield  {journal} {\bibinfo  {journal}
  {Journal of experimental psychology. Human perception and performance}\
  }\textbf {\bibinfo {volume} {10}},\ \bibinfo {pages} {340} (\bibinfo {year}
  {1984})}\BibitemShut {NoStop}%
\bibitem [{\citenamefont {Bybee}(2002)}]{bybee2002}%
  \BibitemOpen
  \bibfield  {author} {\bibinfo {author} {\bibfnamefont {J.}~\bibnamefont
  {Bybee}},\ }\href@noop {} {\bibfield  {journal} {\bibinfo  {journal}
  {Language Variation and Change}\ }\textbf {\bibinfo {volume} {14}},\ \bibinfo
  {pages} {261} (\bibinfo {year} {2002})}\BibitemShut {NoStop}%
\bibitem [{\citenamefont {Havlin}(1995)}]{havlin1995}%
  \BibitemOpen
  \bibfield  {author} {\bibinfo {author} {\bibfnamefont {S.}~\bibnamefont
  {Havlin}},\ }\href@noop {} {\bibfield  {journal} {\bibinfo  {journal}
  {Physica A: Statistical Mechanics and its Applications}\ }\textbf {\bibinfo
  {volume} {216}},\ \bibinfo {pages} {148} (\bibinfo {year}
  {1995})}\BibitemShut {NoStop}%
\bibitem [{\citenamefont {Shannon}(1948)}]{shannon1948}%
  \BibitemOpen
  \bibfield  {author} {\bibinfo {author} {\bibfnamefont {C.~E.}\ \bibnamefont
  {Shannon}},\ }\href@noop {} {\bibfield  {journal} {\bibinfo  {journal} {Bell
  System Technical Journal}\ } (\bibinfo {year} {1948})}\BibitemShut {NoStop}%
\bibitem [{\citenamefont {Shannon}(1951)}]{shannon1951}%
  \BibitemOpen
  \bibfield  {author} {\bibinfo {author} {\bibfnamefont {C.~E.}\ \bibnamefont
  {Shannon}},\ }\href@noop {} {\emph {\bibinfo {title} {Prediction and entropy
  of printed English}}},\ \bibinfo {type} {Tech. Rep.}\ \bibinfo {number} {30}\
  (\bibinfo  {institution} {The Bell System Technical Journal},\ \bibinfo
  {year} {1951})\BibitemShut {NoStop}%
\bibitem [{\citenamefont {Grignetti}(1964)}]{grignetti}%
  \BibitemOpen
  \bibfield  {author} {\bibinfo {author} {\bibfnamefont {M.}~\bibnamefont
  {Grignetti}},\ }\href@noop {} {\bibfield  {journal} {\bibinfo  {journal}
  {Information and Control}\ }\textbf {\bibinfo {volume} {7}},\ \bibinfo
  {pages} {304} (\bibinfo {year} {1964})}\BibitemShut {NoStop}%
\bibitem [{\citenamefont {Piotrovskii}\ \emph {et~al.}(1994)\citenamefont
  {Piotrovskii}, \citenamefont {Pashkovskii},\ and\ \citenamefont
  {Piotrovskii}}]{piotrovskii}%
  \BibitemOpen
  \bibfield  {author} {\bibinfo {author} {\bibfnamefont {R.~G.}\ \bibnamefont
  {Piotrovskii}}, \bibinfo {author} {\bibfnamefont {V.~E.}\ \bibnamefont
  {Pashkovskii}}, \ and\ \bibinfo {author} {\bibfnamefont {V.~R.}\ \bibnamefont
  {Piotrovskii}},\ }\href@noop {} {\bibfield  {journal} {\bibinfo  {journal}
  {Nauchno-Tekhnicheskaya Informatsiya, Seriya 2}\ }\textbf {\bibinfo {volume}
  {28}},\ \bibinfo {pages} {21} (\bibinfo {year} {1994})}\BibitemShut {NoStop}%
\bibitem [{\citenamefont {{McCowan}}\ \emph {et~al.}(1999)\citenamefont
  {{McCowan}}, \citenamefont {Hanser},\ and\ \citenamefont
  {Doyle}}]{mccowan1999}%
  \BibitemOpen
  \bibfield  {author} {\bibinfo {author} {\bibfnamefont {B.}~\bibnamefont
  {{McCowan}}}, \bibinfo {author} {\bibfnamefont {S.~F.}\ \bibnamefont
  {Hanser}}, \ and\ \bibinfo {author} {\bibfnamefont {L.~R.}\ \bibnamefont
  {Doyle}},\ }\href@noop {} {\bibfield  {journal} {\bibinfo  {journal} {Animal
  Behaviour}\ }\textbf {\bibinfo {volume} {57}},\ \bibinfo {pages} {409}
  (\bibinfo {year} {1999})}\BibitemShut {NoStop}%
\bibitem [{\citenamefont {Good}(1953)}]{Good1953}%
  \BibitemOpen
  \bibfield  {author} {\bibinfo {author} {\bibfnamefont {I.~J.}\ \bibnamefont
  {Good}},\ }\href@noop {} {\bibfield  {journal} {\bibinfo  {journal}
  {Biometrika}\ }\textbf {\bibinfo {volume} {40(3-4)}},\ \bibinfo {pages} {237}
  (\bibinfo {year} {1953})}\BibitemShut {NoStop}%
\bibitem [{\citenamefont {Gale}(1994)}]{gale1994}%
  \BibitemOpen
  \bibfield  {author} {\bibinfo {author} {\bibfnamefont {W.}~\bibnamefont
  {Gale}},\ }\href@noop {} {\bibfield  {journal} {\bibinfo  {journal} {Journal
  of Quantitative Linguistics}\ }\textbf {\bibinfo {volume} {2}} (\bibinfo
  {year} {1994})}\BibitemShut {NoStop}%
\bibitem [{\citenamefont {Alexander}\ \emph {et~al.}(1998)\citenamefont
  {Alexander}, \citenamefont {Johnson},\ and\ \citenamefont
  {Weiss}}]{weiss1998}%
  \BibitemOpen
  \bibfield  {author} {\bibinfo {author} {\bibfnamefont {L.}~\bibnamefont
  {Alexander}}, \bibinfo {author} {\bibfnamefont {R.}~\bibnamefont {Johnson}},
  \ and\ \bibinfo {author} {\bibfnamefont {J.}~\bibnamefont {Weiss}},\
  }\href@noop {} {\bibfield  {journal} {\bibinfo  {journal} {Teaching
  Mathematics Applications}\ }\textbf {\bibinfo {volume} {17}},\ \bibinfo
  {pages} {155} (\bibinfo {year} {1998})}\BibitemShut {NoStop}%
\bibitem [{\citenamefont {Samuelsson}(1996)}]{samuelsson1996}%
  \BibitemOpen
  \bibfield  {author} {\bibinfo {author} {\bibfnamefont {C.}~\bibnamefont
  {Samuelsson}},\ }\href@noop {} {\bibfield  {journal} {\bibinfo  {journal}
  {CoRR}\ }\textbf {\bibinfo {volume} {cmp-lg/9606013}} (\bibinfo {year}
  {1996})}\BibitemShut {NoStop}%
\bibitem [{\citenamefont {Cancho}\ and\ \citenamefont
  {Sol\'{e}}(2003)}]{ramon2003}%
  \BibitemOpen
  \bibfield  {author} {\bibinfo {author} {\bibfnamefont {R.~F.}\ \bibnamefont
  {Cancho}}\ and\ \bibinfo {author} {\bibfnamefont {R.~V.}\ \bibnamefont
  {Sol\'{e}}},\ }\href@noop {} {\bibfield  {journal} {\bibinfo  {journal}
  {Proceedings of the National Academy of Sciences of the United States of
  America}\ }\textbf {\bibinfo {volume} {100}},\ \bibinfo {pages} {788}
  (\bibinfo {year} {2003})}\BibitemShut {NoStop}%
\bibitem [{\citenamefont {Mandelbrot}(1953)}]{mandelbrot1953}%
  \BibitemOpen
  \bibfield  {author} {\bibinfo {author} {\bibfnamefont {B.}~\bibnamefont
  {Mandelbrot}},\ }\href@noop {} {\emph {\bibinfo {title} {Contribution {\`a}
  la th{\'e}orie math{\'e}matique des jeux de communication}}},\ Publications
  de l'Institut de Statistique de l'Universit{\'e} de Par{\'\i}s\ (\bibinfo
  {publisher} {Institut Henri Poincar{\'e}},\ \bibinfo {year}
  {1953})\BibitemShut {NoStop}%
\bibitem [{\citenamefont {Mccowan}\ \emph {et~al.}(2005)\citenamefont
  {Mccowan}, \citenamefont {Doyle}, \citenamefont {Jenkins},\ and\
  \citenamefont {Hanser}}]{mccowan2005}%
  \BibitemOpen
  \bibfield  {author} {\bibinfo {author} {\bibfnamefont {B.}~\bibnamefont
  {Mccowan}}, \bibinfo {author} {\bibfnamefont {L.~R.}\ \bibnamefont {Doyle}},
  \bibinfo {author} {\bibfnamefont {J.~M.}\ \bibnamefont {Jenkins}}, \ and\
  \bibinfo {author} {\bibfnamefont {S.~F.}\ \bibnamefont {Hanser}},\
  }\href@noop {} {\bibfield  {journal} {\bibinfo  {journal} {Animal Behaviour}\
  }\textbf {\bibinfo {volume} {69}},\ \bibinfo {pages} {F1} (\bibinfo {year}
  {2005})}\BibitemShut {NoStop}%
\bibitem [{\citenamefont {Mandelbrot}(1965)}]{mandelbrot1965}%
  \BibitemOpen
  \bibfield  {author} {\bibinfo {author} {\bibfnamefont {B.}~\bibnamefont
  {Mandelbrot}},\ }in\ \href@noop {} {\emph {\bibinfo {booktitle} {Scientific
  Psychology: Principles and Approaches}}},\ \bibinfo {editor} {edited by\
  \bibinfo {editor} {\bibfnamefont {B.~B.}\ \bibnamefont {Wolman}}\ and\
  \bibinfo {editor} {\bibfnamefont {E.~N.}\ \bibnamefont {Nagel}}}\ (\bibinfo
  {publisher} {Basic Books},\ \bibinfo {year} {1965})\ pp.\ \bibinfo {pages}
  {550--562}\BibitemShut {NoStop}%
\bibitem [{\citenamefont {Chen}\ and\ \citenamefont {Goodman}(1998)}]{chen98}%
  \BibitemOpen
  \bibfield  {author} {\bibinfo {author} {\bibfnamefont {S.~F.}\ \bibnamefont
  {Chen}}\ and\ \bibinfo {author} {\bibfnamefont {J.}~\bibnamefont {Goodman}},\
  }\href@noop {} {\emph {\bibinfo {title} {An Empirical Study of Smoothing
  Techniques for Language Modeling}}},\ \bibinfo {type} {Tech. Rep.}\ (\bibinfo
   {institution} {Computer Science Group, Harvard University},\ \bibinfo {year}
  {1998})\BibitemShut {NoStop}%
\bibitem [{\citenamefont {Halliday}\ and\ \citenamefont
  {Hasan}(1976)}]{halliday1976}%
  \BibitemOpen
  \bibfield  {author} {\bibinfo {author} {\bibfnamefont {M.~A.~K.}\
  \bibnamefont {Halliday}}\ and\ \bibinfo {author} {\bibfnamefont
  {R.}~\bibnamefont {Hasan}},\ }\href@noop {} {\emph {\bibinfo {title}
  {Cohesion in English}}},\ English language series\ (\bibinfo  {publisher}
  {Longman},\ \bibinfo {year} {1976})\BibitemShut {NoStop}%
\end{thebibliography}%

\end{document}